\documentclass[aps,pra,reprint,superscriptaddress]{revtex4-2}

\usepackage{amsmath}
\usepackage{amsthm}
\usepackage{amssymb}
\usepackage{graphicx}
\usepackage{textcomp}
\usepackage{xcolor}
\usepackage[colorlinks=true, allcolors=blue]{hyperref}
\usepackage{upgreek}
\newcommand{\um}{$\upmu${}m}

\newcommand{\us}{$\upmu${}s}
\newcommand{\uw}{$\upmu${}W}

\renewcommand{\vec}[1]{\ensuremath{\mathbf{#1}}}
\newcommand{\uv}[1]{\ensuremath{\mathbf{\hat{#1}}}} 
\newcommand{\gv}[1]{\ensuremath{\mbox{\boldmath$ #1 $}}} 
\newcommand{\grad}[1]{\gv{\nabla} #1} 

\newcommand{\avg}[1]{\left< #1 \right>}
\renewcommand{\d}[2]{\frac{d #1}{d #2}} 

\begin{document}


\title{Selective ionization of Rydberg atoms to reduce the energy spread of a cold atom focused ion beam}




\author{Kaih T.\ Mitchell}
\email[]{kaih.mitchell@unimelb.edu.au}
\affiliation{School of Physics, University of Melbourne, Parkville, Victoria, 3010 Australia}

\author{Allan Pennings}
\affiliation{School of Physics, University of Melbourne, Parkville, Victoria, 3010 Australia}

\author{Rory W.\ Speirs}
\affiliation{School of Physics, University of Melbourne, Parkville, Victoria, 3010 Australia}

\author{Christopher J.\ Billington}
\affiliation{School of Physics, University of Melbourne, Parkville, Victoria, 3010 Australia}

\author{Andrew J.\ McCulloch}
\affiliation{School of Physics, University of Melbourne, Parkville, Victoria, 3010 Australia}
\affiliation{School of Natural Sciences, University of Tasmania, Hobart, Tasmania 7001, Australia}

\author{Alexander A.\ Wood}
\affiliation{School of Physics, University of Melbourne, Parkville, Victoria, 3010 Australia}

\author{Robert E.\ Scholten}
\affiliation{School of Physics, University of Melbourne, Parkville, Victoria, 3010 Australia}


\date{\today}

\begin{abstract}
The energy spread of a focused ion beam causes chromatic aberration that limits the focal spot size and resolution for imaging and fabrication. Ion beams based on photoionization of neutral atoms can have a much smaller energy spread and higher brightness than conventional liquid metal ion sources and are thus capable of higher resolution. We present a method for using selective ionization of Rydberg atoms to reduce the beam energy spread in a cold atom focused ion beam. We produce experimental maps of the ionization rate of $^{85}$Rb near the classical ionization threshold, predict the energy spread of an ion beam produced from these states and demonstrate energy spread reduction \textit{in situ} in a focused ion beam. We use a novel method to measure the energy spread, using only components present in many commercial FIB systems. Selecting different states changed the energy spread by up to 50\%, with opportunities to further reduce the energy spread by constructing a more favorable electric field gradient and finding atomic states with better ionization characteristics.
\end{abstract}


\maketitle

\section{Introduction}
Ion beam sources based on photoionization of cold atoms are capable of achieving higher brightness, and thus higher resolution at the same current, than conventional ion sources~\cite{mcclelland2016}. As smaller spot sizes are achieved, the effect of chromatic aberration becomes increasingly significant, requiring a reduction in the beam energy spread. Cold atom ion sources are capable of sophisticated ionization schemes, such as selective ionization for reducing the energy spread~\cite{kime2013, moufarej2017, mcculloch2017rydberg, hahn_thesis, hahn2021}, and heralding to produce single ions or provide active beam trajectory correction~\cite{mcculloch2017herald, sahin2017, lopez2019, trimeche2020}. The conventional liquid metal ion source (LMIS)~\cite{hagen2008, bauerdick2013} does not allow for fine control over the ionization process, so cannot use these techniques for reducing the energy spread.

Significant reduction of the energy spread can be achieved by employing a \textit{field ionization} scheme, where an external electric field modifies the ionization properties of the atom. Several studies discuss the possibility of using field ionization of Rydberg states of alkali atoms to reduce the energy spread of cold-atom electron and ion beams~\cite{kime2013, moufarej2017, mcculloch2017rydberg}. This work directly demonstrates the application of this scheme to reduce the energy spread \textit{in situ} in a focused ion beam apparatus.

Cold-atom focused ion beams (CFIBs) use a cold atom ion source, and have previously been demonstrated using lithium~\cite{knuffman2011}, cesium~\cite{steele2017, viteau2016, steele2021} and rubidium~\cite{tenhaafthesis, tenhaaf2017, xu2023, li2024}. Preliminary investigations into using field ionization to reduce the beam energy spread of electrons and ions~\cite{kime2013, moufarej2017, mcculloch2017rydberg} did not integrate the ion source into a functional focused ion beam (FIB) system, unlike in this work.

\begin{figure*}[tb]
\begin{center}
\includegraphics[width=\textwidth]{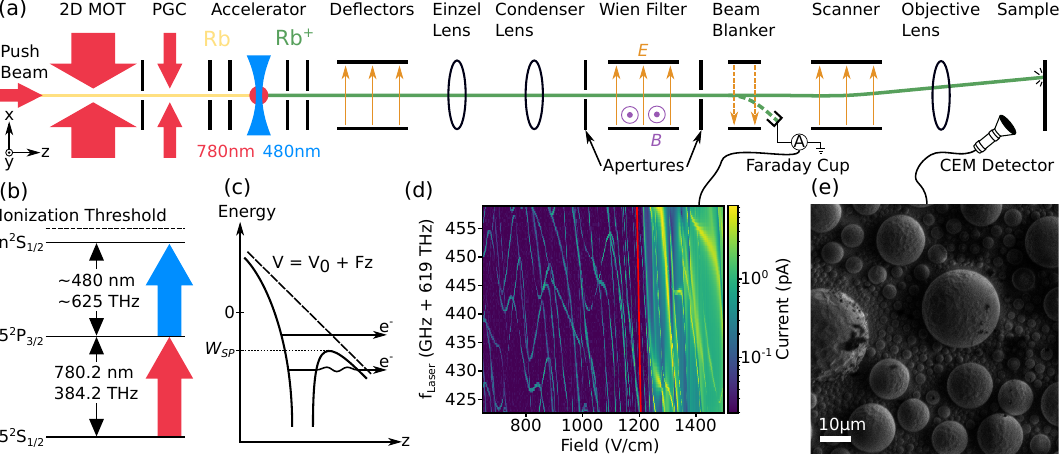}
\caption{(a) Schematic diagram of the CFIB system. (b) Energy level diagram showing the 780\,nm excitation and 480\,nm ionization transitions. (c) Sketch of an atomic potential in an external electric field. The dashed line represents the potential of the external electric field along the $z$ axis of the FIB, which increases in gradient as the electric field strength $F$ is increased. The ionization energy of the atom can be reduced by applying an external field and below-threshold ionization can occur. $W_{SP}$ is the saddle point energy and represents the classical ionization threshold. (d) Color plot of ion current as a function of electric field strength and ionization laser frequency. Ion current was measured by switching on the beam blanker to direct the ion beam into the Faraday cup. The red line is the classical ionization threshold at energy $W_{SP}$. (e) Rubidium ion micrograph of tin spheres on carbon, acquired at 3.0\,keV beam energy by scanning the ion beam across the sample and collecting secondary electrons with the CEM detector.}
\label{fig:one}
\end{center}
\end{figure*}

\section{Experimental apparatus}\label{sec:experimental details}
Our CFIB apparatus~\cite{mitchell_thesis} is shown schematically in Figure\;\ref{fig:one}(a). The ion source consists of a two-dimensional mag\-neto-optical trap (2D MOT), followed by transverse atomic beam cooling, photoionization and electrostatic acceleration~\cite{mcclelland2016,steele2017,viteau2016,tenhaaf2017}. The ion source is attached by a gate valve and a custom adapter to a commercial FIB column (Orsay Physics FIB40~\footnote{Orsay Physics, \url{www.orsayphysics.com}}), replacing the conventional gallium LMIS.

Rydberg excitation was performed by two perpendicular laser beams: the excitation (780\,nm, MOGLabs ECD) and ionization (480\,nm, Toptica TA-SHG Pro) lasers. The atomic transitions driven by the two laser beams are shown schematically in Figure\;\ref{fig:one}(b). The excitation laser couples $^{85}$Rb atoms from the $5^2S_{1/2}\;F=3$ to the $5^2P_{3/2}\;F'=4$ state, and the ionization laser then excited the atoms to a Rydberg state from which they were subsequently field ionized. Typical excitation beam power was 20\,\uw{} and typical ionization power was 500\,mW. A 2\,\uw{} repump laser beam was also combined with the excitation beam to prevent optical depumping of the excitation transition into the $5^2S_{1/2}\;F=2$ dark state, by exciting out of this state to the $5^2P_{3/2}\;F'=3$ state. The laser beams were focused by in-vacuum lenses to a spot size of 16\,\um. Four electrodes surround the ionization region and produced the electric field used for accelerating the ions and inducing the Stark shift in the atomic states. Figure\;\ref{fig:two}(a) shows a schematic of the ionization region. Photoionization occurs in the region between the second and third electrode. \vfill

\section{Using Field ionization to Reduce the Energy Spread}\label{sec:field_ionization_theory}

In the presence of an electric field, the energy of atomic states is shifted by the Stark effect~\cite{gallagher1994, cleanthes1990}. At strong electric fields, the Stark shift becomes highly non-linear and results in states that are highly sensitive to electric field strength, and therefore the ionization rate changes rapidly with changes in electric field strength for a given state energy. Using this electric field sensitivity could reduce the energy spread below what could be achieved with direct ionization. We show that exploiting states with an ionization rate that changes rapidly with electric field can be used to control the ionization probability and reduce the energy spread of the ion beam.

\begin{figure}[tb]
\begin{center}
\includegraphics[width=\columnwidth]{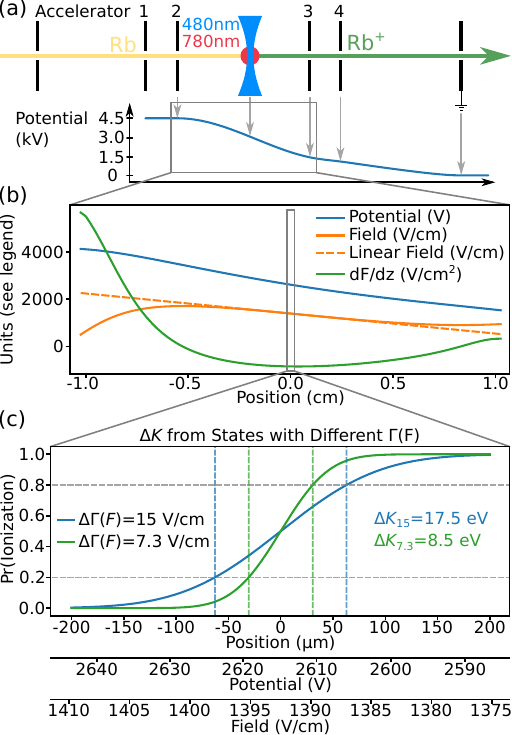}
\caption{(a) Sketch of the accelerator assembly, showing typical electrostatic potential configuration. Electrodes 2 and 3 are separated by 2.05\,cm. Diagram is not to scale. (b) Plots of the accelerator potential, electric field and electric field gradient $\left(\d{F}{z}\right)$ along the central axis of the ionization region, as determined by SIMION calculations. The accelerator voltages were those used in the experiments described in Sec.\,\ref{subsec:wien_stark} (4433\,V, 4433\,V, 1572\,V, 1253\,V). The potential at the center of the ionization region is 2616\,V, which sets the beam energy. A linear expansion of the field around $z=0$ illustrates that the field is linear in distance close to the center of the ionization region. The plot is bounded by the accelerator electrodes at $\pm$1.025\,cm. (c) Plot of Eq.\,\ref{eqn:ionization_theory} using Gaussian $\Gamma(F)$ distributions with two different FWHM values that were measured from experimental Stark maps. Dashed vertical lines show $z_{20}$ and $z_{80}$ for the two different states. The beam energy spread is reduced by choosing states with narrower $\Gamma(F)$. The values of $V(z)$ and $F(z)$ were calculated using SIMION.}
\label{fig:two}
\end{center}
\end{figure}

Following the treatment in Ref.\,\cite{moufarej2017}, we assume that an atom is moving along the $z$ direction at a constant velocity, $v$, through an electric field oriented along the $z$ direction, $F(z) = \d{F}{z} z + F(0)$. Figure\;\ref{fig:two}(b) shows calculated values of potential, field and field gradient in the ionization region for an electrode potential configuration similar to that used in our experiments. The cumulative probability of ionizing an atom vs position is given by~\cite{moufarej2017}
\begin{equation}
P[F(z)] = 1 - \exp\left[-\frac{1}{v\d{F}{z}}\int_{F(-\infty)}^{F(z)}\Gamma(F')dF'\right]\,,
\label{eqn:ionization_theory}
\end{equation}
where $\Gamma(F)$ is the ionization rate at a given field $F$.

Critically, for states where the ionization rate varies rapidly with respect to $F$, i.e.\ $\Gamma(F)$ has a narrow distribution, the probability of ionization is tightly confined in $z$, and consequently a small range of potentials, giving a narrow beam energy spread. Finding states with $\Gamma(F)$ that is narrow with respect to $F$ is crucial to this work. 

From Eq.\,\ref{eqn:ionization_theory} and data on the potential in the ionization region (Fig.\,\ref{fig:two}(b)), we can predict the beam energy spread for states with different $\Gamma(F)$. Figure\;\ref{fig:two}(c) plots the ionization probability distributions for two different states and determines the resulting energy spread. When calculating the ionization probability, the dominant ionization mechanism is assumed to be field ionization, i.e.\ we assume the atoms have a sufficiently long excited state lifetime to remain excited and ionize outside of the intersecting laser beam volume. We define the range over which atoms are between 20\% and 80\% likely to ionize as $\Delta z_{20,80}$. From this range we can calculate the beam energy spread as
\begin{equation}
\Delta K \equiv e\left[V(z_{80}) - V(z_{20})\right] \approx eF(0) \Delta z_{20,80}\,, \label{eqn:delta_k_2080}
\end{equation}
where $e=\text{1.602}\times \text{10}^{-19}\text{\,C}$ is the elementary charge, $V(z)$ is the potential at position $z$, and $F(0)$ is the electric field at the center of the ionization region. The plots shown in Figure\;\ref{fig:two}(c) are calculations of the expected energy spread for two different $\Gamma(F)$ distributions with full width at half maximum (FWHM) of 15\,V/cm and 7.3\,V/cm. These plots predict the outcome of the experiment performed in Sec.\,\ref{sec:beam_energy_spread}, where the beam energy spread is measured. The values for $F(z)$ and $\d{F}{z}$ were determined from SIMION calculations~\cite{simion} (Fig.\,\ref{fig:two}(b)). Note that the ionization probability was normalized by dividing by the probability as $z\rightarrow\infty$.

The significant influence of the Stark effect on ionization rates is due to the modified atomic potential changing the electron binding~\cite{gallagher1994, cleanthes1990}. Figure\;\ref{fig:one}(c) shows a sketch of an atomic potential with an external electric field applied. The external field creates a saddle point below the field-free ionization threshold, i.e. electrons can escape from the atom with less energy than in zero electric field. This \textit{saddle point energy}, $W_{SP}$, also referred to as the \textit{classical ionization threshold}, is given by~\cite{gallagher1994}
\begin{equation}
W_{SP} = -2\sqrt{\frac{e^3 F}{4\pi\varepsilon_0}}\,,
\end{equation}
where $e$ is the elementary charge, $F$ is the external electric field strength and $\varepsilon_0$ is the vacuum permittivity. As the electric field strength is increased, the difference between $W_{SP}$ and the field-free ionization threshold increases, allowing atoms to ionize more readily. Quantum tunneling through classical potential barriers allows atoms to ionize without exceeding the classical ionization threshold. This effect is also illustrated in Figure\;\ref{fig:one}(c), as a state with energy below $W_{SP}$ is able to ionize, and is referred to as ionizing \textit{below threshold}. Much of the ionization performed in this work is done close to the classical ionization threshold.

\section{Experimental Stark Mapping to Find Low-Energy-Spread States}\label{sec:stark_mapping_experiments}

To experimentally determine the optimal combinations of ionization laser frequency and electric field strength, we measured the ionization rate as a function of these parameters, a process known as \textit{Stark mapping}~\cite{kime2013, moufarej2017, mcculloch2017rydberg, hahn_thesis, hahn2021}. To create a Stark map, as shown in Figure\;\ref{fig:one}(d), we scanned the ionization laser frequency over its full modehop-free scan range (30\,GHz) while measuring its frequency using a wavemeter (High Finesse WS-6). While scanning the laser, the ion current was measured using a Faraday cup and transimpedance amplifier (Femto DDPCPA-300) to produce a spectrum of beam current vs laser frequency. The ion current was measured with an amplifier gain of 10$^{11}$\,V/A and a bandwidth of 20\,Hz, and spectra were acquired with 100 laser frequency steps using an acquisition time of 100\,ms/step. Each time the accelerator potentials were changed, a fast, wide field-of-view, low-resolution scan was performed, using the CEM to collect secondary electrons from the sample, to ensure the beam was properly aligned. The laser frequency was measured at the end points of a number of short frequency sweep segments and interpolated for intermediate values, as the frequency measurement could not be reliably synchronized with the beam current data acquisition. We repeated this procedure for each value of electric field strength, $F$, in the accelerator region to create a Stark map. The nominal electric fields were calculated using the potentials on accelerator electrodes 2 and 3 and their separation distance. Calculations~\cite{moufarej2017, harmin1984, comparatPrivate} of the expected Stark maps, and SIMION calculations of the fields, suggest a 1-3\% discrepancy (dependent on curvature of electric field, and thus beam energy) between the nominally predicted electric field and the true field. This discrepancy was not found to be of any consequence.

\section{Measuring the Beam Energy Spread}\label{sec:beam_energy_spread}

The most direct method for measuring the beam energy spread is with a purpose-built beam energy spread analyzer~\cite{tenhaaf2018}, but this additional equipment is not feasible in many FIB systems. Another method would be to measure the spot size of the ion beam while changing the ionization state. However, systematics may prevent the spot size from being optimized sufficiently to be confident it was limited by chromatic aberration. We developed a novel solution using a Wien filter in the commercial FIB column (Fig.\,\ref{fig:one}(a)) to selectively blur the images along one axis in a manner proportional to the beam energy spread in that axis.  An advantage of using the Wien filter in this way is that it does not require the ion beam spot size to be perfectly optimized, and therefore limited by chromatic aberration, to observe the energy spread. Wien filters are commonly found in conventional FIB systems for selecting isotopes or elements in a liquid metal alloy ion source~\cite{bauerdick2013}. A Wien filter can also be used as a monochromator, but typical implementations do not have sufficient resolution to directly reduce the energy spread.  

\subsection{Wien Blurring as a Diagnostic}\label{subsec:wien_blur_theory}
The Wien filter is a velocity-selective filter, using crossed electric and magnetic fields, which is found in many commercial FIB systems. In typical usage, the Wien filter is used to separate different isotopes in an LMIS or liquid metal alloy ion source~\cite{bischoff2016, pilz2019, fit4nano_roadmap2023}. As ion beams have an inherent velocity spread, passing a beam through the Wien filter changes the transverse velocity of each particle, depending on its kinetic energy. Using a beam that passes through the Wien filter to image a sample results in an image that is blurred along one axis by an amount proportional to the energy spread of the beam.

We assume that the ion beam has an energy spread given by a function of $K$, the kinetic energy. We modeled the energy spread as a Gaussian function with a mean of $K_0$ and an 20-80 width $\Delta K$~\cite{moufarej2017}, as in Eq.\,\ref{eqn:delta_k_2080}. When the beam passes through the filter, the energy spread function will be transformed into a transverse velocity function, related to the amount of deflection. We assume that the electric field of the Wien filter ($E$, not to be confused with $F$, the electric field in the ionization region) is oriented in the positive $x$ direction ($\vec{E} = E \uv{x}$), the magnetic field is oriented in the positive $y$ direction ($\vec{B} = B\uv{y}$) and the beam travels with velocity $\vec{v} = v_z\uv{z}$ along the positive $z$ direction. The acceleration applied by a Wien filter with a given electric and magnetic field strength on particles with mass $m$ and charge $q$ is given by~\cite{orloff2009}
\begin{equation}
\vec{a}(v_z) = \frac{q}{m}\left(E-v_zB\right)\uv{x}\,.
\end{equation}
Substituting in the kinetic energy, $K=\frac{1}{2}m{v_z}^2$, gives the acceleration as a function of beam energy as
\begin{equation}
\vec{a}(K) = \frac{q}{m}\left(E-\sqrt{\frac{2K}{m}}B\right)\uv{x}\,.
\end{equation}
We expect the beam energy spread will only be a few eV compared to the beam energy of several keV, so we use a first-order Taylor expansion around the point $K_0$, i.e.\ where $v_z=\frac{E}{B}$, to find 
\begin{equation}
\vec{a}(K)\approx \frac{qE}{2m}\left(1-\frac{K}{K_0}\right)\uv{x}\,.
\end{equation} 
This linear relationship between kinetic energy and the acceleration imparted by the Wien filter shows that, in the limit that the energy spread is much smaller than the mean energy ($\Delta K \ll K_0$), the Wien filter will impart a transverse velocity distribution on the beam that is representative of the beam energy distribution. If the Wien filter $B$ and $E$ fields are not correctly set so that ions with energy $K_0$ are undeflected, the beam will receive an asymmetrical shift, resulting in the same spread, plus a net deflection. 

For ions traveling with longitudinal speed $v_z = \sqrt{\frac{2K}{m}}$ through an active region of length $L$, the time spent interacting with the filter is $t = \frac{L}{v_z}$. Assuming an ion has $v_x=0$ and energy $K$, 
\begin{equation}
\vec{v}_x(K) = \vec{a}(K)t \approx \frac{qLE}{2\sqrt{2K_0m}}\left(1-\frac{K}{K_0}\right)\uv{x}\,,
\end{equation}
which gives a linear mapping from the kinetic energy distribution to the transverse velocity distribution imparted by the Wien filter. For a beam with an energy spread $\Delta K$, the resulting velocity spread in the $x$ direction will be
\begin{equation}\label{eqn:wien_delta_v}
\Delta v_x = \frac{qLE}{2\sqrt{2K_0m}}\frac{\Delta K}{K_0}\,.
\end{equation}
From this we can infer that to maximize $\Delta v_x$, so that the blurring effect is most obvious, we must maximize the electric field strength in the Wien filter and set the magnetic field strength to avoid unwanted deflection. 

\begin{figure}[tb]
\begin{center}
\includegraphics[width=\columnwidth]{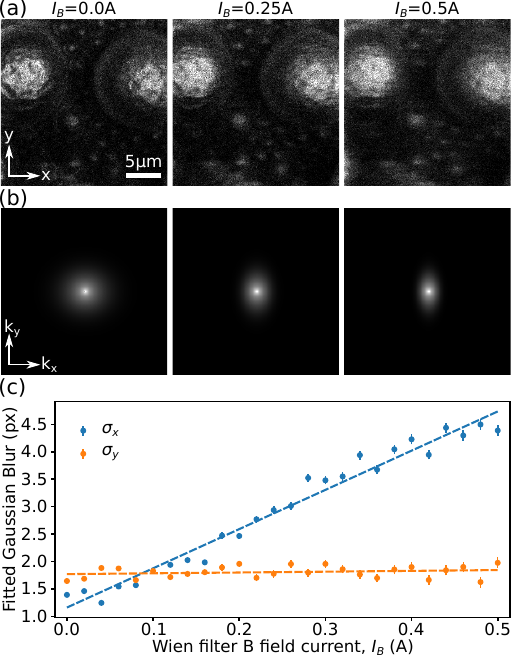}
\caption{(a) Examples of rubidium ion micrographs of tin spheres on carbon acquired at different values of the Wien filter strength ($I_B$ as shown, and appropriate $E$ field voltage, here 14.6\,V at 0.5\,A). As the Wien filter strength increases, the amount of blur in the horizontal direction increases. This effect is most evident when inspecting the smaller tin spheres. (b) Examples of the natural logarithm of the fitted Fourier transform for the corresponding images in (a). $k_x$ and $k_y$ represent spatial frequencies in the $x$ and $y$ directions. As the images become more blurred in the $x$ direction, the width of the feature in the Fourier transform becomes narrower in the $k_x$ direction. The values of $\sigma_x$ and $\sigma_y$ are extracted from these fits (Eq.\,\ref{eqn:2D_FFT_fit}). (c) The fitted Gaussian blur values ($\sigma$) for the $x$ and $y$ axes of a series of images, showing that as the Wien filter $B$ field current, $I_B$, is increased, $\sigma_x$ increases linearly while $\sigma_y$ remains constant, indicating that blur increases only along the $x$ axis. Dashed lines are fits. Error bars show the fitting error.}
\label{fig:three}
\end{center}
\end{figure}

Additionally, the effect of the Wien filter is dependent on the ion spatial distribution at the position along the beam that the filter is applied. This effect can be understood by examining how the Wien filter changes the emittance of the ion beam. The normalized emittance of a particle beam is given by~\cite{reiser2008}
\begin{equation}\label{eqn:emittance_ch5}
\epsilon_x = \sqrt{K} \sqrt{\avg{x^2}\avg{x'^2} - \avg{xx'}^2}\,,
\end{equation}
where $x$ is the transverse position and $x'$ is the angle from the $z$ axis of a particle in the beam. At the focus $\epsilon_x \propto \sigma_x$. By making the paraxial approximation, the angle the beam makes to the $z$ axis is proportional to the transverse velocity, since for a given particle
\begin{equation}\label{eqn:x'_of_vx}
x' = \frac{v_x}{v_z} = v_x\sqrt{\frac{m}{2K}}\,.
\end{equation}
The Wien filter will affect the emittance by changing the distribution of $v_x$ and hence $x'$. The effect of the Wien filter can be considered as an additional transverse velocity, $w$, drawn from a distribution symmetric around zero, and not correlated to $x$ or $x'$. Substituting Eq.\,\ref{eqn:x'_of_vx} into Eq.\,\ref{eqn:emittance_ch5} and $v_x \rightarrow v_x + w$ gives
\begin{equation}
\epsilon_x = \sqrt{\frac{m}{2}}\sqrt{\avg{x^2}\avg{{v_x}^2} +\avg{x^2}\avg{w^2} - \avg{xv_x}^2}\,,
\end{equation}
where we have assumed that $x$ and $v_x$ are independent of $w$, i.e.\ $\avg{xw} = \avg{v_x w} = 0$. Thus the emittance increases when $\avg{w^2}$ is non-zero. If the ion beam has a narrow waist inside the Wien filter, i.e.\ $\avg{x^2}\approx 0$, due to focusing from the condenser lens, then the effect of the Wien filter will be minimal compared to a collimated beam with the same emittance. A strongly converging ion beam already has a large transverse velocity spread, so the additional velocity imparted by the Wien filter is less significant. For this reason, at a given beam energy, the condenser lens voltage was adjusted to find the value that gave the most blurring when the Wien filter was applied. The objective lens was also adjusted to maintain image focus. 

\subsection{Experimental Verification of Wien Filter Blurring}\label{subsec:wien_verification}

We parameterized the effect of the Wien filter by the driving current of the $B$ field coils, $I_B$, which ranged from 0\,A to 1.0\,A, corresponding to $B$ field of 62\,mT. For a given $B$ field, we tuned and fixed the $E$ field such that there was no image translation when the filter was on or off. Small amounts of image translation caused by a net deflection of the beam by the Wien filter should not affect results, as the image analysis algorithm (Appendix\;\ref{sec:image_analysis}) only measures the blur and ignores translations of the image. 

To measure the beam energy spread, a series of images were acquired with different values for the Wien filter strength. These images were acquired in a random sequence to ensure any increase in the image blur was attributable to the Wien filter, and not caused by gradual changes in FIB parameters. To quantitatively assess the increase in image blur as the Wien filter was applied, we used an image analysis technique that involved taking the Fourier transform of the image and fitting a function to extract the amount of blur. The function fitted was the convolution of a 2D Gaussian function and a function representing the Fourier transform of a step change in image intensity. The analysis technique is discussed in detail in Appendix\;\ref{sec:image_analysis}. Figure\;\ref{fig:three} shows a subset of a sequence of images and a plot of the estimated blur values vs $I_B$. These images, the corresponding Fourier transforms and the plot of fitted blur values vs $I_B$ show that blur increases along the $x$ axis, but not the $y$ axis, thus validating that the Wien filter can induce a monoaxial blurring. A maximum $I_B$ value of 0.5\,A was used as a reliable magnetic field calibration could not be achieved at higher Wien filter currents. 

To calibrate the energy spread measurement, we adjusted the beam energy by a known amount from $K_0$ and measured the image position shift. Two images were acquired at each beam energy, one with the Wien filter on and one with it off. The two images were correlated to find the shift in the $x$ direction. To avoid changing the electric field strength in the ionization region, $F$, when the beam energy was adjusted, SIMION calculations were performed to find accelerator voltages that minimize the change in $F$ when changing $K$. The experimentally obtained calibration factor, for the configuration used in Sec.\,\ref{subsec:wien_stark}, was $\eta=\text{2.02}\pm\text{0.25\,eV/px}$ at $I_B=\text{0.5\,A}$.

\subsection{Demonstrating Reduced Energy Spread}\label{subsec:wien_stark}

\begin{figure}[tb]
\begin{center}
\includegraphics[width=\columnwidth]{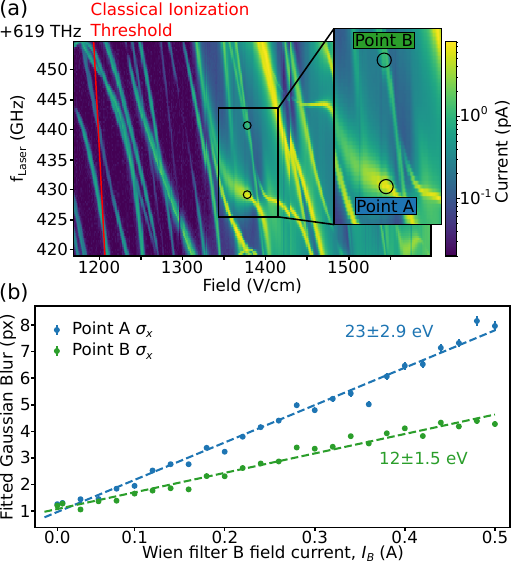}
\caption{(a) Stark map showing two points with $\Gamma(F)$ of different widths that were used to demonstrate the effect of ionization volume on the beam energy spread. Inset shows an enlarged view of the region of interest. (b) Plots of $\sigma_x$ vs $I_B$ for the two points shown in the Stark map. Dashed lines are fits. A steeper gradient indicates a larger energy spread. Point B has a narrower $\Gamma(F)$ which results in a two times smaller beam energy spread than at Point A.}
\label{fig:four}
\end{center}
\end{figure}

We used the Wien filter method to demonstrate that our ionization scheme can target specific atomic states and reduce the beam energy spread. This experiment was performed by choosing an electric field strength and atomic transition, and comparing the blurring with the Wien filter technique for different transitions. Stark maps indicated which transitions would produce a narrower energy spread, $\Delta K$, which is related to the width of the transitions along the $F$ axis on a Stark map, $\Gamma(F)$. To minimize the number of experimental variables, we used a constant accelerator potential for measurements at both points A and B, so that the nominal beam energy did not change. We also selected above-threshold states so that they were strongly ionizing and data could be acquired rapidly, though this was purely for convenience. Figure\;\ref{fig:four} shows the results of the energy spread reduction experiment. 

A field value of 1393\,V/cm allowed two strongly ionizing states with different $\Gamma(F)$ to be ionized while only changing the laser frequency within a single mode-hop-free tuning range. From SIMION calculations, the resulting field gradient was -850\,V/cm$^2$ (Fig.\,\ref{fig:two}(b)). The laser frequency (619.435\,THz, exciting to principle quantum number $n \approx 26$ at ${F=0\text{V/cm}}$) was selected to be close to the classical ionization threshold at this field. Two points with different values of $\Gamma(F)$ were chosen to demonstrate a reduction in the beam energy spread, with FWHM of 15\,V/cm FWHM (point A, see Figure\;\ref{fig:four}(a)) and 7.3\,V/cm FWHM (point B). The standard deviation of the Gaussian blur, $\sigma_x$, at $I_B=\text{0.5\,A}$ was used to calculate $\Delta K$, using
\begin{equation}
\Delta K \approx 1.683\,\sigma_x\,\eta\,,
\end{equation}
where $\eta=\text{2.02}\pm\text{0.25\,eV/px}$ at $I_B=\text{0.5\,A}$, and the factor of 1.683 converts from standard deviation to 20-80 width. The measured energy spread values were 23$\pm$2.9\,eV for point A and 12$\pm$1.5\,eV for point B. These results are compared to values estimated using Eq.\,\ref{eqn:ionization_theory} and Eq.\,\ref{eqn:delta_k_2080} and shown in Table\;\ref{tab:wien_results}. The predicted relative reduction in energy spread agrees to within error margins. The experimentally determined values were larger than the predicted values by 5.8\,eV and 3.6\,eV respectively, suggesting that the calibration factor $\eta$ may have been overestimated, or there are additional systematic errors that have not been accounted for. Drift in the beam alignment caused by ions charging insulating surfaces in the ionization region prevented full optimization of the FIB apparatus.

In the accelerator configuration used here, the ionization region defined by $\Delta z_{20,80}$ at points A and B (61\,\um{} and 126\,\um{}) in Figure\;\ref{fig:two}(c), is larger than the volume defined by the intersecting excitation and ionization lasers (16\,\um{}). If ionization only occurred in the intersecting laser volume, the energy spread difference between ionizing from point A or B would be less than 3\%, rather than the 50\% difference observed. This result indicates that the atoms are being excited to a Rydberg state and subsequently being field-ionized after leaving the intersecting laser volume, even though the energy of the states are nominally above threshold. As Rydberg states can have relatively long lifetimes, of order 10\,\us{}~\cite{branden2009}, there is a possibility that atoms will remain in a Rydberg state and propagate several millimeters before being ionized. Thus, changing the ionization state could change the ionization energy by several hundred eV, depending on propagation distance. In practice, we observed that changing the ionization state did not significantly change the propagation distance. If ionization occurred in a significantly different location, the ions would have very different kinetic energy, and the Wien filter would no longer be correctly tuned, causing a lateral shift in the image position, which was not observed. The lack of observable lateral shift means that the change to beam energy was insignificant compared to the change to the energy spread. 

\begin{table}[tbp]
\centering
\begin{tabular}{crrr}
\hline
Point & $\Gamma(F)$ FWHM & \begin{tabular}{r}Estimated\\$\Delta K$\end{tabular} & \begin{tabular}{r}Experimental\\$\Delta K$\end{tabular}\\\hline
A & 15.0$\pm$0.5\,V/cm & 18$\pm$0.6\,eV & 23$\pm$2.9\,eV\\
B & 7.3$\pm$0.5\,V/cm & 8.5$\pm$0.6\,eV & 12$\pm$1.5\,eV\\
\hline
A/B & 2.05$\pm$0.15 & 2.1$\pm$0.21 & 1.9$\pm$0.48\\\hline
\end{tabular}
\caption{Results from the energy spread reduction experiment. Estimated values were calculated using the methods detailed in Sec.\,\ref{sec:field_ionization_theory}}
\label{tab:wien_results}
\end{table}

Both the measured and predicted energy spread at point B are still worse than the 4 to 5\,eV typical for a gallium LMIS~\cite{mcclelland2016, hagen2008}. Despite not out-performing gallium in this instance, the demonstration of the technique of using selective ionization to reduce the energy spread by 50\% is promising. An accelerator assembly that can produce low fields and high field gradients, as well as selecting states with narrow $\Gamma(F)$ will further reduce the energy spread, as predicted by Eq.\,\ref{eqn:ionization_theory}. In the case of exceptionally low fields and comparatively high field gradients (10\,V/cm, 100\,V/cm$^2$), theory predicts selective field ionization could produce a $\Delta K$ of 3\,meV~\cite{kime2013}, which is 1000 times smaller than a gallium LMIS and will allow operation at lower beam energies and with higher brightness.

The influence of the beam energy spread on the focal spot size is dependent on several variables including beam energy and objective lens focal length. In general, the spot size is minimized when the contributions from emittance and chromatic and spherical aberrations are balanced, so over-optimizing one parameter eventually has diminishing returns~\cite{barth1996,wang1991}. For typical FIB parameters chromatic aberration is insignificant once $\frac{\Delta K}{K} < 10^{-5}$, i.e.\ 0.3\,eV for a 30\,keV beam~\cite{mcclelland2016, barth1996}. At lower beam energies, and higher brightness, $\Delta K$ needs to be reduced further. 

\section{Discussion}
The demonstration described in this work shows that selective ionization from specific states can reduce the energy spread of the focused ion beam. Further work needs to be done to address technical problems with our accelerator assembly, including slow response to changing potential and ongoing drift, which prevented long-term stable operation at high-resolution. Additionally, the FIB requires improvement to demonstrate significant advantages relative to typical LMIS performance, including by increasing beam current and energy, but these aspects have already been demonstrated elsewhere~\cite{steele2017, viteau2016, xu2022, li2024}. Experimental and theoretical searches for states with the necessary high ionization rate and narrow $\Gamma(F)$ would improve performance and may also identify better atomic species or isotopes for state-selective ionization. Using below-threshold states should provide some advantages for reducing the energy spread, as these states are observed to be narrower than above-threshold states, at the cost of ionizing less readily. An additional avenue for achieving narrower energy spreads is to ionize from so-called ``exceptional" Rydberg states. Such states undergo interference narrowing, and transition from weakly ionizing to rapidly ionizing, and back again, over a small range of electric field strength values~\cite{kime2013, mcculloch2017rydberg, moufarej2017}. Finding exceptional states, either via an experimental or theoretical search, could further reduce the beam energy spread. Aside from $\Gamma(F)$, varying the combination of ionization laser energy, $F(0)$ and $\d{F}{z}$ can be used to minimize the ultimate beam energy spread~\cite{kime2013}. Choosing an optimal electric field strength and gradient configuration is a tradeoff between achieving a narrow energy spread and producing a high-brightness beam. For example, a low field means a low $\d{V}{z}$, resulting in a narrow energy spread, but the lower fields come at the cost of transverse broadening. High $\d{F}{z}$ can drastically reduce the energy spread, but Gauss's law $\left(\grad\cdot\textbf{F}=0\right)$~\cite{griffiths1999} requires that for an electric field to have a gradient in the $z$ direction, there must also be a gradient in the transverse directions, leading to an undesirable lensing effect. 

Higher beam energies require higher field strengths in the ionization region to minimize the effect of transverse acceleration of the ions towards the grounded outer casing of the ionization region, meaning that the minimum achievable energy spread is dependent on beam energy. Further, increasing the intensity of the ionization laser, either by increasing the laser power, a more tightly confined beam, or implementing an in-vacuum optical build-up cavity~\cite{tenhaafthesis}, could additionally increase the ionization efficiency.

\section{Conclusion}

We have demonstrated selective ionization using Rydberg states in a rubidium CFIB. Adjusting the electric field and ionization laser frequency allowed us to manipulate the states that atoms are excited to prior to ionization, and to control how the atoms ultimately ionize. We developed a technique to quantitatively measure the beam energy spread by intentionally inducing image blur with a Wien filter, avoiding the need for the spot size to be chromatic aberration limited or additional analysis equipment. We demonstrated that changing the parameters used for ionizing the atoms could reduce the energy spread of the resulting beam by a factor of two. Demonstrating that selective field ionization can be used \textit{in situ} in a FIB system for reducing the beam energy spread highlights the advanced capabilities of CFIB systems. The minimum beam energy spread we measured was 12.1$\pm$1.5\,eV, which is not better than a gallium LMIS (4 to 5\,eV), but the significance of this work is demonstrating that the energy spread can be reduced by selective ionization in a CFIB, something that is impossible with an LMIS. 

\begin{acknowledgments}
The authors thank F.J.\ Robicheaux and D.\ Comparat for providing code for simulating Stark maps. This work was supported by the Australian Research Council (DP200103452). K.T.M.\ was supported by an Australian Government Research Training Program (RTP) Scholarship. A.A.W.\ was supported by an ARC DECRA Fellowship (DE210101093).
\end{acknowledgments}

\appendix
\section{Image Analysis}\label{sec:image_analysis}

Various techniques and algorithms for quantifying the blur of SEM images exist in the literature~\cite{joy2000, ishitani2002_hann, babin2006, brostrom2022}. We modify the SIRAF (Spatial Image Resolution Assessment by Fourier analysis) algorithm~\cite{brostrom2022} which takes the 2D Fourier transform of an image and fits a function that extracts a value for the equivalent Gaussian blur. This analysis technique requires no user input or subjective judgments, either of image quality or the Fourier transform. The SIRAF algorithm required modification to measure differences in the resolution between the $x$ and $y$ axes.

The SIRAF algorithm assumes that a blurred image of a sharp feature can be modeled as a step function $h(x)$, convolved with a Gaussian $g(x)$. The convolution of an image with a Gaussian function is a common technique for introducing blurring~\cite{steck_optics}. By determining $g(x)$ we are effectively determining the point-spread function of the imaging system~\cite{hecht1987}.

\begin{figure}[tbp]
\begin{center}
\includegraphics[width=1.0\columnwidth]{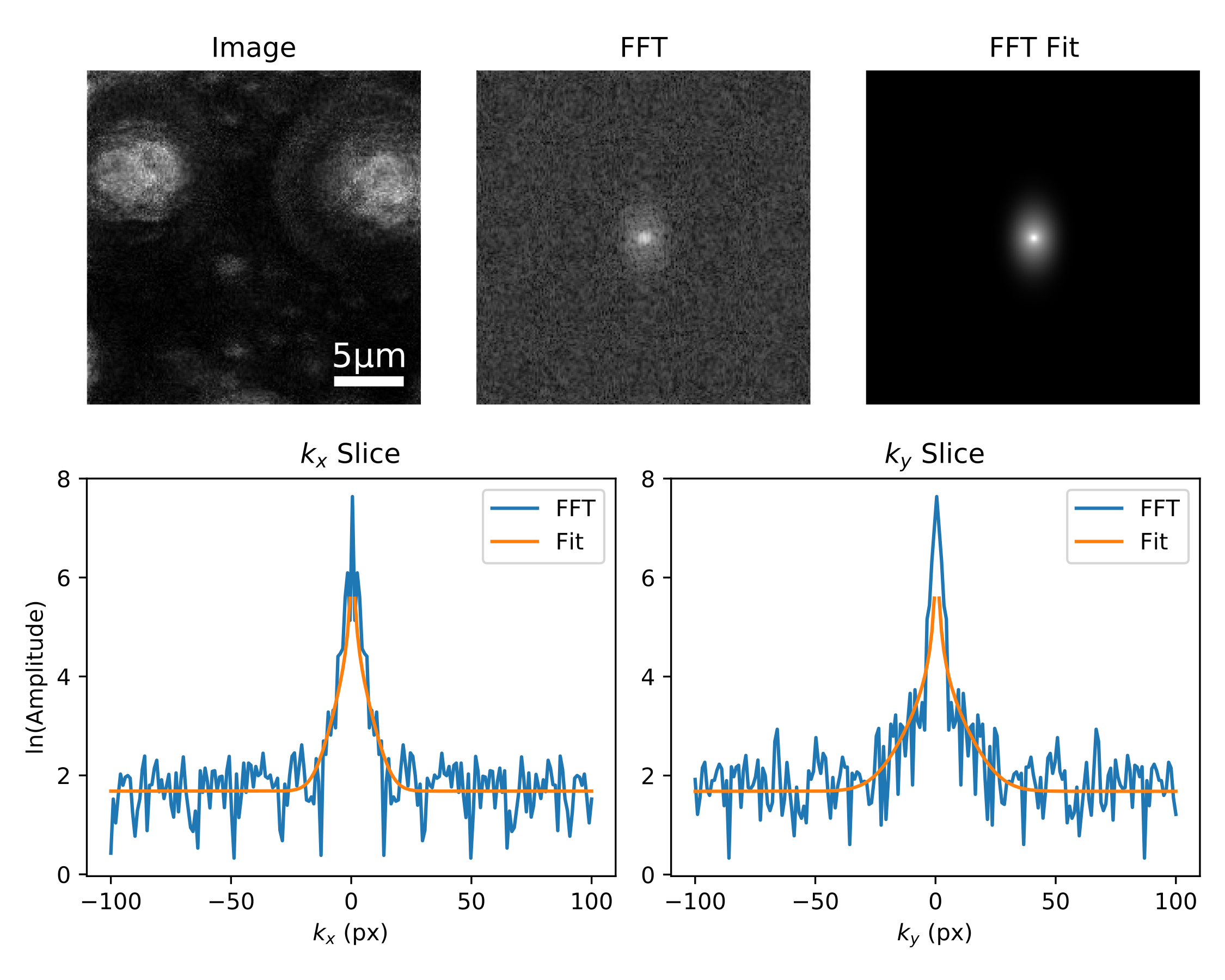}
\caption{Plots demonstrating our algorithm for determining a value for the image blur in two dimensions. Top left: Original image of tin spheres on carbon. Top middle: Grayscale map of amplitude of the 2D Fourier transform of the image, plotted with a log scale. Top right: Fit of Eq.\,\ref{eqn:2D_FFT_fit} to the 2D Fourier transform, plotted on a log scale. Bottom: Lineouts along the $k_x$ and $k_y$ axes of the Fourier transform and the fits to show agreement. The original image was taken when the Wien filter was being used to deliberately blur the image in the $x$ direction. The effect of this is apparent in the Fourier transform as the spatial frequencies in the $k_x$ axis are reduced, making the ellipse appear stretched in the vertical direction. $f(k_x,k_y)$ is undefined at $(k_x,k_y)=(0,0)$, so the fit does not include this point.}
\label{fig:FFT_fit_plots}
\end{center}
\end{figure} 

To estimate the amount of blur separately for each axis, a more general 2D form of the SIRAF algorithm was required. We define an elliptical 2D Gaussian, centered on the origin and rotated by angle $\theta$ as
\begin{equation}
g(x,y) = A \exp\left[-\left(ax^2 + 2bxy + cy^2\right)\right]\,,\label{eqn:2D_gaussian}
\end{equation}
where
\begin{eqnarray}
\quad a = & \frac{\cos^2\theta}{2{\sigma_i}^2} + \frac{\sin^2\theta}{2{\sigma_j}^2}\,, \\
b = & -\frac{\sin(2\theta)}{4{\sigma_i}^2} + \frac{\sin(2\theta)}{4{\sigma_j}^2}\,, \\
c = & \frac{\sin^2\theta}{2{\sigma_i}^2} + \frac{\cos^2\theta}{2{\sigma_j}^2}\,,
\end{eqnarray}
and $\sigma_i$ and $\sigma_j$ are the standard deviations of the Gaussian profiles along the semiaxes of the ellipse rotated by angle $\theta$. When $\theta=0$, $\sigma_i = \sigma_x$ and $\sigma_j = \sigma_y$.
The Fourier transform of $g(x,y)$ is
\begin{equation}
G(k_x, k_y) = \frac{2\pi A \exp\left[\frac{4\pi^2\left(c{k_x}^2 - 2bk_x k_y + a{k_y}^2\right)}{4b^2-4ac}\right]}{\sqrt{-4b^2 + 4ac}}\,.
\end{equation}
The SIRAF algorithm uses the following definition of $h(x)$ to represent an edge
\begin{equation}
h(x) = \frac{1}{2}\left(1 + \text{sgn}(x)\right)\,,
\end{equation} 
where $x$ is a coordinate in image space and $\text{sgn}(x)$ is the sign function. The Fourier transform of $h(x)$ is
\begin{equation}
H(k) = \frac{1}{2}\left(\delta(k) - \frac{i}{\pi k}\right)\,.
\label{eqn:fft_of_step}
\end{equation}
Ignoring the Dirac delta, as it contains no spatial frequency information at $k=0$, and taking the absolute value gives
\begin{equation}
\left|H(k)\right| = \frac{1}{2\pi\sqrt{k^2}} \,,
\end{equation}
which we generalize to 2D as 
\begin{equation}
\left|H(k_x, k_y)\right| = \frac{1}{2\pi\sqrt{{k_x}^2 + {k_y}^2}}\,.
\end{equation}
We then perform the convolution of $h(x,y)$ and $g(x,y)$ using the convolution theorem~\cite{goodman2005}, and take the absolute value to give
\begin{equation}
\left|H(k_x, k_y)G(k_x, k_y)\right| = \frac{ A \exp\left[\frac{4\pi^2\left(c{k_x}^2 - 2bk_x k_y + a{k_y}^2\right)}{4b^2-4ac}\right]}{\sqrt{{k_x}^2 + {k_y}^2}\sqrt{-4b^2 + 4ac}}\,.
\end{equation}
To extract values for $\sigma_x$ and $\sigma_y$ from an image, we apply a Hann window, then Fourier transform the image. As the fast Fourier transform (FFT) algorithm uses periodic boundary conditions, the Hann window tapers pixel intensities to zero at the edges of the image to prevent boundary discontinuities from affecting the resulting spectrum~\cite{ishitani2002_hann}. To make fitting easier, a fixed offset of $1$ is added to the absolute value of the Fourier transformed image and the natural logarithm is taken. We then use a fitting function 
\begin{equation}
f(k_x, k_y) = c + \ln\left[1 + \left|H(k_x, k_y)G(k_x, k_y)\right| \right]\,.
\label{eqn:2D_FFT_fit}
\end{equation}
From the fit of Eq.\,\ref{eqn:2D_FFT_fit}, the $\sigma_x$ and $\sigma_y$ parameters can be extracted, giving values for the blur in the two axes of the image. An example of a rubidium FIB image being analyzed with this technique is shown in Figure\;\ref{fig:FFT_fit_plots}.

We tested this method on images that had been artificially blurred by a known amount to verify that the correct values for $\sigma_x$ and $\sigma_y$ were returned to within fitting error. 

\vfill

\bibliography{bibliography}

\end{document}